\documentclass[pra,tightenlines,showpacs,nofootinbib,groupedaddress]{revtex4}
\usepackage{dcolumn,graphicx}
\begin{document}
\preprint{UNR August 2001-\today }
\title{ Reevaluation of the role of nuclear
uncertainties in experiments on atomic parity violation with isotopic chains}

\author{  Andrei Derevianko }
\email{andrei@unr.edu}
\affiliation { Department of Physics, University of Nevada, Reno,
Nevada 89557}
\author{Sergey G.~Porsev}
%\email{porsev@physics.unr.edu},
\altaffiliation{Permanent Address: Petersburg Nuclear Physics
Institute, Gatchina, Leningrad district, 188300, Russia.}
\affiliation { Department of Physics, University of Nevada, Reno,
Nevada 89557}
%\affiliation{Petersburg Nuclear Physics Institute, Gatchina,
%Leningrad district, 188300, Russia.}
\date{\today}

\begin{abstract}
In light of new data on neutron distributions from experiments
with antiprotonic atoms [ Trzcinska {\it et al.}, Phys. Rev. Lett.
87, 082501 (2001)], we reexamine the role of nuclear-structure
uncertainties in the interpretation of measurements of parity
violation in atoms using chains of isotopes of the same element.
With these new nuclear data, we find an improvement in the
sensitivity of isotopic chain measurements to ``new physics''
beyond the standard model. We compare possible constraints on
``new physics'' with the most accurate to date single-isotope
probe of parity violation in the Cs atom. We conclude that
presently isotopic chain experiments employing atoms with nuclear
charges $Z \lesssim 50$ may result in more accurate tests of the
weak interaction.
\end{abstract}


\maketitle

\section{Introduction}
Atomic parity non-conservation~\cite{Khr91,BouBou97} (PNC)
provides powerful constraints on extensions to the standard model
of elementary particles in the low-energy electroweak sector. For
example, a deviation in the observed weak charge of an atomic
nucleus, $Q_W$, from the prediction of the standard model may hint
at the existence of extra neutral-gauge $Z$-boson. Other possible
``new physics'' scenarios are discussed, e.g., in
Ref.~\cite{GroAguAms00}.

The most accurate to date measurement of atomic PNC has been
carried out by Wieman and co-workers~\cite{WooBenCho97,BenWie99}
using a single isotope of atomic cesium, $^{133}$Cs. In such
measurements one determines a parity-violating signal $E_{\rm
PNC}$, related to the weak charge as $E_{\rm PNC}= k_{\rm PNC}\,
Q_W$. The parameter $k_{\rm PNC}$ is supplied  from sophisticated
atomic-structure calculations. Even for the relatively
well-understood univalent Cs atom, the accuracy of the calculation
of $k_{\rm PNC}$ remains the limiting factor in the determination
of the weak charge.

There is an ongoing discussion in the
literature~\cite{BenWie99,Der00,Der01a,DzuHarJoh01,KozPorTup01,JohBedSof01,MilSus02,DzuFlaGin02}
about whether the $^{133}$Cs weak charge deviates from the
prediction of the standard model. This possible
deviation may be interpreted as an indication for an extra
neutral-gauge boson $Z'$~\cite{CasCurDom99,Ros00}; there
were numerous discussions in the literature about
implications of the possible deviation.
 It is clear that independent
tests of parity violation in atoms are required at least at the
level of the present 1\% accuracy for Cs. It is worth mentioning
that PNC measurements were also carried out in Tl
\cite{VetMeeMaj95,EdwPhiBai95}, Pb \cite{MeeVetMaj95}, and Bi
(see, e.~g., \cite{WarThoSta93}). Among these,  the simplest atom
is Tl, but even for Tl the theoretical uncertainty for $k_{\rm
PNC}$ is a factor of a few larger than that for
Cs~\cite{KozPorJoh01,DzuFlaSil87J}.

An alternative approach allowing one to circumvent the
difficulties of atomic-structure calculations was proposed
by~\citet{DzuFlaKhr86}. The main idea was to form a ratio
$\mathcal{R}$ of PNC amplitudes for two isotopes of the same
element, thus cancelling out associated uncertainties of the
atomic theory. However, \citet{ForPanWil90} pointed out a
conceptual limitation of this method -- an enhanced sensitivity of
possible constraints on ``new physics'' to uncertainties in the
neutron distributions. As an example, the differences between
neutron and proton root-mean-square radii for $^{133}$Cs differ by
a factor of four in relativistic and non-relativistic
nuclear-structure calculations and depend on nuclear models.
Unfortunately, at the present level of theoretical understanding
of neutron distributions, such large nuclear-structure
uncertainties would preclude an extraction of useful information
on weak interactions from isotopic ratios measured for heavy
atoms.

Given the inadequate accuracy of nuclear-structure calculations for
the analysis of PNC measurements based on isotopic ratios,
here we investigate the role of uncertainties in neutron
distributions using {\em empirical} data. Recently,
\citet{TrzJasLub01} deduced differences between root-mean-square
radii $R_n$ and $R_p$  of neutron and proton distributions from
experiments with antiprotonic atoms. A wide range of stable nuclei
were investigated and the differences were approximated by a linear
dependence suggested in Ref.~\cite{PetRav96}
%------------------------------------------------------------------
\begin{eqnarray}
\Delta R_{np} = (-0.04 \pm 0.03) + (1.01 \pm 0.15)\frac{N-Z}{N+Z} \, \,  \textrm{fm} .
\label{Rnp1}
\end{eqnarray}
%------------------------------------------------------------------
Here $\Delta R_{np} = R_n - R_p$, $Z$ is the nuclear charge, and
$N$ is the number of neutrons. Recently, this result was employed
to estimate the nuclear-structure uncertainty for parity-violating
amplitude in Cs~\cite{Der01a}.
In light of the new nuclear data we
reexamine the suitability of isotopic chain measurements for
studies of parity violation in atoms. We find that the
nuclear-structure uncertainty in possible probes of ``new
physics'' with isotopic chains is reduced by the new
antiprotonic-atom data. We compare constraints on the direct ``new
physics'' with what is currently the most accurate single-isotope
probe of parity violation in $^{133}$Cs. We conclude that
presently isotopic chain experiments with atoms having $Z \lesssim
50$ may be competitive with this single-isotope determination.

\section{Background}
In a typical atomic PNC setup, one considers a transition between
two atomic states $|i \rangle$ and $|f \rangle$ of the same
nominal parity. The weak interaction admixes the states $|
n\rangle$ of the opposite parity,  leading to the otherwise
forbidden parity-violating amplitude
%------------------------------------------------------------------
\begin{eqnarray}
   E_{\rm PNC} &=&  \sum_{n} \left[
\frac{\langle f | D_z | n  \rangle
      \langle n | H_{\rm W} | i \rangle}{E_i - E_n}\right.
\nonumber \\                                                    % (1)
      &+&
\left.\frac{\langle f | H_{\rm W} | n  \rangle
      \langle n | D_z | i \rangle}{E_f - E_n} \right],
\label{EPNC}
\end{eqnarray}
%------------------------------------------------------------------
where $D$ is the electric-dipole operator and $H_{\rm W}$ is the
Hamiltonian of the electron-nucleus weak interaction. As
demonstrated by \citet{PolForWil92}, matrix elements  of $H_{\rm
W}$ may be represented as
%------------------------------------------------------------------
\begin{eqnarray}
\langle j |H_W| i \rangle =                                  % (2)
\frac{G_F}{2 \sqrt{2}} \ C_{ji} \ R_{p}^{2 \gamma-2} \ Q_W (N,Z) \, ,
\label{HW}
\end{eqnarray}
%------------------------------------------------------------------
where factor $C_{ji}$ depends on atomic wavefunctions
and  $\gamma = \sqrt{1 - (\alpha Z)^2}$.

Including the dependence on nuclear shapes, the nuclear weak charge $Q_W (N,Z)$
may be represented at the tree level as
%------------------------------------------------------------------
\begin{eqnarray}
 Q_W =                                                    % (3)
-N \, q_n + Z \, q_p \ (1 - 4 \,{\rm sin}^2 \theta_W) + \Delta Q_{\rm new}.
\label{QW}
\end{eqnarray}
%------------------------------------------------------------------
Here $\sin^2 \theta_W = 0.23117\,(16)$~\cite{GroAguAms00} and
quantities $q_n$ and $q_p$, introduced in \cite{ForPanWil90}, depend on neutron and proton
distributions inside a nucleus. It should be noted that quantities $q_n$ and $q_p$ are
numerically very close to one.
For example, in the ``sharp edge'' model
of nuclear density distribution~\cite{ForPanWil90}
%------------------------------------------------------------------
\begin{eqnarray}
q_n = 1-\frac{3}{70} \left( \alpha Z \right)^2
     \left[1+5 \left( \frac{R_n}{R_p} \right)^{2} \right] \,.        % (11)
\label{qn}
\end{eqnarray}
%------------------------------------------------------------------
More sophisticated expressions may be found in Ref.~\cite{JamSan99},
but the accuracy of the above formula is sufficient for the goals of the present work.
We omitted radiative corrections in the definition of the weak
charge, Eq.~(\ref{QW}). These contributions are important in the studies
of ``oblique'' corrections, discussed, e.g., in Ref.~\cite{PolForWil92,Ros96}.
Here, motivated by possible deviation of the Cs weak charge from the
prediction of standard model, we analyze constraints on direct tree-level
``new physics''.
The term $\Delta Q_{\rm new}$ in Eq.~(\ref{QW}) characterizes ``new
physics''. Following~\citet{Ram99}, we represent it as a combination
of couplings to up ($u$) and down ($d$) quarks, i.e.
%------------------------------------------------------------------
\begin{eqnarray}
 \Delta Q_{\rm new} &=& (2Z + N) \, h_u + (Z + 2N) \, h_d \\ \nonumber
     & \equiv & Z \ h_p + N \ h_n,                       % (4)
\label{Qnew}
\end{eqnarray}
%------------------------------------------------------------------
where $h_p = 2 h_u + h_d$ and $h_n = 2 h_d + h_u$ are couplings to
protons and neutrons. Various elementary-particle scenarios for
these interactions were reviewed in Ref.~\cite{Ram99}.
Finally,
%------------------------------------------------------------------
\begin{eqnarray}
 Q_W = N \ h_0 + Z \ h_p + N \ h_n,                       % (5)
\label{Qh0}
\end{eqnarray}
%------------------------------------------------------------------
with
%------------------------------------------------------------------
\begin{eqnarray}
 h_0 \equiv -q_n + \frac{Z}{N} \ q_p \ (1 - 4 \, {\rm sin}^2 \theta_W)
      \approx -q_n \, .                                         % (6)
\label{h0}
\end{eqnarray}
%------------------------------------------------------------------

As an outcome of the analysis of PNC experiments one would like to set bounds
on ``new physics'' couplings $h_p, h_n$ or equivalently $h_u, h_d$; below
we summarize relevant analysis from Refs.~\cite{ForPanWil90,PolForWil92,Ram99}.
The PNC amplitudes $E_{PNC}$ and $E'_{PNC}$  are measured for two isotopes
of the same element with
neutron numbers $N$ and $N'=N+\Delta N$, and the ratio is formed
%------------------------------------------------------------------
\begin{eqnarray}
\mathcal{R} = \frac{E_{PNC}}{E_{PNC}^{\prime}}              % (7)
 =  \frac{Q_W}{Q_W^\prime}
    \left( \frac{R_p}{R_p^\prime} \right)^{2\gamma-2}.
\label{R}
\end{eqnarray}
%------------------------------------------------------------------
Here all quantities with primes are for the isotope with
$N^\prime$ neutrons. Using Eq.~(\ref{Qh0})  one obtains
%------------------------------------------------------------------
\begin{eqnarray}
\mathcal{R} =
\mathcal{R}_0 \left\{ 1 + \frac{Z\, \Delta
N}{N\, N^\prime} \, \frac{h_p}{h_0} + \left( \frac{Z}{N^\prime}\,
h_p + h_n \right) \frac{h_0^\prime - h_0}{h_0^\prime \, h_0}
\right\} \label{RR0}
\end{eqnarray}
%------------------------------------------------------------------
with $\mathcal{R}_{0} \equiv \left( \frac{R_p}{R_p^\prime} \right)^{2\gamma-2}
\frac{N\, h_0}{N^\prime \, h_0^\prime}$. The last term in the above expression
may be safely neglected and we  determine a contribution of ``new
physics''
%------------------------------------------------------------------
\begin{eqnarray}
\mathcal{F}=\frac{ h_p}{h_0}= \left(  \frac{\mathcal{R}}{\mathcal{R}_0}-1 \right)
\frac{N\, N^\prime}{Z\, \Delta N}.                                  % (9)
\label{F}
\end{eqnarray}
%------------------------------------------------------------------
In the absence of new couplings $\mathcal{F}=0$.
It may seem counterintuitive that the isotopic ratios  are
sensitive to the new physics encapsulated in couplings to protons ($h_p$) instead
of those to neutrons ($h_n$).
The dependence on $h_p$ may be easily demonstrated with an alternative ratio $(Q_W/N - Q'_W/N')/(Q_W/N + Q'_W/N')$;
this ratio is straightforwardly reduced to $Z \Delta N /(2 N N') h_p$.

The constraints on $h_p$, Eq.(~\ref{F}), are affected by (i) the experimental error bar in
$\mathcal{R}$ and (ii) uncertainties in $\mathcal{R}_0$ which are
induced by insufficient knowledge of nuclear distributions.
Explicitly,
%------------------------------------------------------------------
\begin{eqnarray}
\delta \mathcal{F}= \frac{N\, N^\prime}{Z\, \Delta N}
\left\{\frac{\delta \mathcal{R}}{\mathcal{R}_0}
       + \delta (\Delta q_n) \right\}.                            % (10)
\label{delF}
\end{eqnarray}
%------------------------------------------------------------------
Here $\Delta q_n \equiv q_n - q_n^\prime$ and we assumed
$R_p \approx R_p^\prime$. The radii of
proton distributions are known with sufficient accuracy~\cite{FriBerHei95}
and we  disregarded associated uncertainties.
Finally,
%------------------------------------------------------------------
\begin{eqnarray}
\delta \mathcal{F}= \frac{N\, N^\prime}{Z\, \Delta N}
\left\{\frac{\delta \mathcal{R}}{\mathcal{R}_0}
     + \frac{3}{7} \left( \alpha Z \right)^2                    % (12)
       \frac{\delta \Delta R_n}{R_p} \right\} \, ,
\label{delFR}
\end{eqnarray}
%------------------------------------------------------------------
with $ \Delta R_n = R'_n -R_n$. The above expression is similar to
the results of \cite{PolForWil92}.

\section{Results and discussion}
We assume below that in Eq.~(\ref{delFR}) the experimental
errors $\delta \mathcal{R}$ may be neglected in comparison to
nuclear-structure uncertainty. By contrast to the previous
discussions~\cite{ForPanWil90,PolForWil92,Ram99}
of atomic parity-violation
in isotope chains we employ the empirical Eq.~(\ref{Rnp1}) to estimate radii of
neutron distributions; this relation was deduced from experiments
with antiprotonic atoms~\cite{TrzJasLub01}.
To estimate the error bar in the differences
$\Delta R_{np}-\Delta R'_{np}$, we formed all possible isotope pairs from
the original 21 point data set of Ref.~\cite{TrzJasLub01} and obtained with
the least-square method
$\Delta R'_{np} - \Delta R_{np} = (0 \pm 0.003)+ (1.01 \pm 0.04)
\left\{
(N'-Z')/(N'+Z') - (N-Z)/(N+Z)\right\}$ fm. Instead of
a single-parameter fit, we performed a two-parameter
fit because there is no strong theoretical reason to believe that the
difference $\Delta R'_{np} - \Delta R_{np}$ should vanish
for two distinct nuclei with the same relative neutron excess,
$(N'-Z')/(N'+Z') = (N-Z)/(N+Z)$.
%In Eq.~(\ref{Eqn_err}), employed in our subsequent analysis,
Such obtained (statistical) error bars are very small.
However, given insufficient information on systematic errors in \cite{TrzJasLub01},
in our subsequent analysis we  retained more conservative
uncertainties from Eq.~(\ref{Rnp1}).
Based on error bars in Eq.~(\ref{Rnp1})
we set
\begin{equation}
\delta \Delta R_n \approx \delta \left(\Delta R'_{np} - \Delta R_{np} \right ) \approx
\left[ (0.03)^2 +
\left\{0.15
\left(\frac{N'-Z}{N'+Z} - \frac{N-Z}{N+Z}\right)
\right\}^2
\right]^{1/2} \,
\mathrm{fm} \, .
\label{Eqn_err}
\end{equation}
The first (isotope-independent) term in this expression dominates for $Z>20$; for
heavy atoms $ \delta \Delta R_n \approx 0.03$ fm.
It is worth emphasizing that the Eq.~(\ref{Rnp1}) for differences between
neutron and proton r.m.s. radii was obtained in Ref.~\cite{TrzJasLub01}
with data for {\em stable} isotopes; it is expected that nonstable
isotopes may exhibit anomalous $\Delta R_{np}$.

We require
the nuclear-structure uncertainty in $\delta \mathcal{F}$ be lower
than the current limits deduced from the most accurate to date single-isotope $^{133}$Cs
determination. The single-isotope measurements are sensitive to a different
combination of new $h_u$ and $h_d$, $u-e$ and $d-e$ couplings.
For illustration we parameterize
$h_u = \lambda h_d$. For example, $\lambda=0$ arises in
analyses of extra neutral-gauge $Z$ bosons in E$_6$ theories and $\lambda=1$
corresponds to pure isoscalar couplings~\cite{Ram99}. We obtain
\begin{equation}
\delta \mathcal{F} ( ^{133}\mathrm{Cs}) = \frac{\delta h_p}{h_0} \approx
\frac{\delta Q_W}{Q_W}
\frac{N}{Z + \frac{2 + \lambda}{2 \lambda +1} N} \, .
\end{equation}
We set $\frac{\delta Q_W}{Q_W} \approx 0.01$, i.e. the present 1\%
precision of the determination of the weak charge in $^{133}$Cs, and
find
\begin{equation}
\delta \mathcal{F} ( ^{133}\mathrm{Cs}) = \left\{
\begin{array}{l@{\quad,\quad}l@{\quad}l}
     3.7 \times 10^{-3}  &   \lambda =0 & h_u=0 \\
     5.9 \times 10^{-3}  &   \lambda =1 & h_u=h_d\\
     8.3 \times 10^{-3}  &   \lambda =\pm \infty& h_d=0\\
     \infty              &   \lambda = - \frac{Z+2N}{2Z+N} & h_u \approx -1.22 h_d
 \end{array} \right.  \, .
\label{Eqn_CsConstraints}
\end{equation}
In our illustrative example, the single-isotope bounds set on
``new physics''  encapsulated in $h_p$ are clearly
model-dependent. We note that the single-isotope  $^{133}$Cs
measurement is insensitive to a particular scenario $h_u \approx
-1.22 h_d$, which may be directly probed by the measurements with
chains of isotopes or constrained by other electroweak
observables.

Given an experimental precision $\delta \mathcal{R}/\mathcal{R}$
in determination of  PNC amplitudes, the uncertainty (\ref{delFR})
may be minimized by using a pair of  isotopes with the maximum
possible spread of neutron numbers $\Delta N$. Based on
Eq.~(\ref{Eqn_err}) and (\ref{delFR}) we calculated $\delta
\mathcal{F}$ for such stable isotope pairs for Ba, Sm, Yb, and Pb.
We have chosen these atoms mostly because PNC experiments were
carried out for them, or at least discussed in the
literature~\cite{For97,KuwEndFuk99,LucWarSta98,Kim01,Bud98,MeeVetMaj93}.
>From the results compiled in Table~\ref{Tab_iso}, it is clear that
the present nuclear-structure uncertainty
still may cloud a competitive extraction of ``new physics'' from isotopic chain experiments
for these atoms.
Compared to single-isotope $^{133}$Cs determination, measurements
with isotopes of Ba and Sm would be two times less sensitive
to extra neutral-gauge $Z$-bosons and would have a comparable sensitivity to
new isoscalar physics (see Eq.~(\ref{Eqn_CsConstraints}).)
Possible constraints from heavier Yb and Pb would be affected by the nuclear uncertainty
to a larger extent.

%===================================================================
\begin{table}
\caption{Contribution of nuclear-structure uncertainty to a constraint
on ``new physics'' $\delta \mathcal{F}$
for representative isotope pairs.}
\label{Tab_iso}
\begin{ruledtabular}
\begin{tabular}{cccc}
\multicolumn{1}{c}{Atom} & \multicolumn{2}{c}{Mass numbers $A$} &
\multicolumn{1}{c}{$\delta \mathcal{F}\times 10^3$} \\
\hline
Ba $(Z=56)$ & 130 & 138 & 6.2 \\
Sm $(Z=62)$ & 144 & 154 & 6.5      \\
Yb $(Z=70)$ & 168 & 176 & 12 \\
Pb $(Z=82)$ & 204 & 208 & 39 \\
\hline
$^{133}$Cs $(Z=55)$\footnotemark[1]
 & 133 & -- & 3.7\\
$^{133}$Cs $(Z=55)$\footnotemark[2]
 & 133 & -- & 5.9\\
\end{tabular}
\end{ruledtabular}
\footnotetext[1]{ Single-isotope constraint for extra
neutral-gauge $Z$-bosons scenario,
Eq.~(\protect\ref{Eqn_CsConstraints}). } %
\footnotetext[2]{ Single-isotope constraint for isoscalar scenario,
Eq.~(\protect\ref{Eqn_CsConstraints}). } %

\end{table}
%===================================================================

\begin{figure}
\centerline{\includegraphics*[scale=0.5]{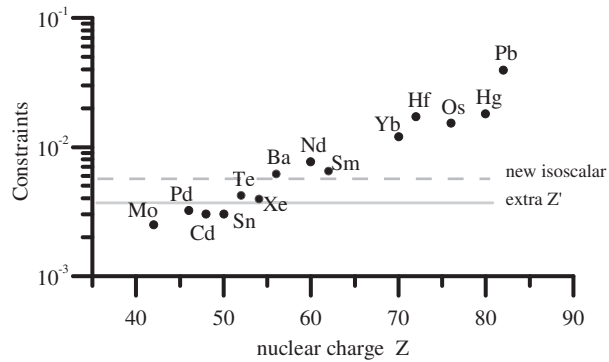}}
\caption{ Contribution of nuclear-structure uncertainty to possible constraints
on ``new physics'' $\delta \mathcal{F}$
for atoms with nuclear charges $40<Z \leq 82$. Horizontal lines represent
limits derived from single-isotope $^{133}$Cs PNC analysis in the
isoscalar (dashed line) and extra neutral-gauge boson $Z'$ (solid line) scenarios.
   \label{Fig_manyAtoms} }
\end{figure}

Now we proceed with a search for atoms suitable for isotopic ratio
experiments given present nuclear-structure uncertainties. In
Fig.~\ref{Fig_manyAtoms} we summarize results for atoms with
nuclear charges $40<Z \leq 82$. To minimize the effect of
experimental error $\delta \mathcal R$ in  $\delta \mathcal{F}$,
the spread in neutron numbers $\Delta N$ should be chosen as large
as possible; we only considered atoms with stable isotopes so that
$\Delta N \ge 8$ ( $\Delta N=4$ for Pb). We approximate $N \approx
1.5\, Z$, $R_p \propto Z^{1/3}$, and the error $\delta \Delta R_n
\approx 0.03$ fm. Thus the nuclear-structure uncertainty in the
determination of ``new physics'' $\delta \mathcal{F}$ grows as
$Z^{8/3}$, explaining a general trend in Fig.~\ref{Fig_manyAtoms}.
We compare the uncertainties to constraints set by the Cs
determination (horizontal lines).  We conclude that the isotopic
chain measurements in atoms with nuclear charges $Z \lesssim 50$
may provide comparable limits on couplings for the interesting
extra $Z$ scenario. For these elements an interpretation of the
measurements in terms of  direct new physics may be relatively
free of nuclear-structure uncertainties. It is worth emphasizing
that extra $Z'$ were discussed recently in connection with
a possible deviation of $^{133}$Cs weak charge from the prediction
of the standard model.

We would like to briefly comment on the required experimental
accuracy in determination of ratio $\mathcal{R}$ of the
parity-violating amplitudes. Approximating $N \approx 1.5 \, Z$ we
find
\[
\frac{\delta \mathcal{R}}{\mathcal{R}} \lesssim
0.4 \frac{\Delta N}{Z} \delta \mathcal{F}
\]
We set $\delta \mathcal{F}$ to constraints derived from the
determination of $^{133}$Cs weak charge, Eq.~(\ref{Eqn_CsConstraints}).
We arrive at
\begin{equation}
\frac{\delta \mathcal{R}}{\mathcal{R}_0} \lesssim
0.4 \frac{\Delta N}{Z} \delta \mathcal{F}( ^{133}\mathrm{Cs})
\approx 0.02 \, \, \frac{\Delta N}{Z} \, .
\label{Eqn_exptErr}
\end{equation}
The required accuracy in the ratio of PNC amplitudes $\mathcal{R}$
is in the order of 0.3\% for Ba and Sm,
0.2\% for Yb, and 0.1\% for Pb. The required experimental
error is less demanding  for lighter atoms.

So far the most accurate measurement of parity-violating amplitude
was carried out in Cs~\cite{WooBenCho97}; the achieved accuracy
was 0.35\%. As first noted by \citet{BouBou75}, the matrix
elements of the weak interaction scale as $Z^3$; the
parity-violating amplitude may be weaker for atoms with nuclear
charges $Z \lesssim 50$ which are lighter than Cs ($Z=55$).
However, the required experimental error in ratios of PNC
amplitudes, Eq.~(\ref{Eqn_exptErr}), is less demanding  for
lighter atoms. Also an  enhancement of  PNC amplitude may arise
due to an admixture to the initial/final atomic state of an
energetically close intermediate state of an opposite parity by
the weak interaction. For example, calculations
\cite{DeM95,PorRakKoz95} demonstrated that the PNC amplitude for
the $6s^2 \, ^1S_0 \rightarrow 5d\,6s \, ^3D_1$ transition in Yb
is approximately 100 times larger than in Cs.

We conclude that at the present level of understanding of neutron
distributions, atoms with nuclear charges $Z \lesssim 50$ may be
suitable for competitive tests of parity violation with isotopic
ratios. If parity-violating enhancement scenarios would be
realized for such atoms, the experiments may become feasible. It
is worth carrying out a systematic search for enhanced PNC
amplitudes for atoms and ions with $Z \lesssim 50$. Such an
atomic-structure search is certainly a nontrivial task, requiring
in most of the cases an accurate account of correlations. For
example, \citet{XiaMouYou90} argued that the PNC amplitude for the
$6s^2 \, ^1\!S_0 \rightarrow 5d7s \, {}^3\!D_1$ transition in Ba
is an order of magnitude larger than in Cs. Their semi-empirical
calculation was based on assumption that an intermediate state
$6s7p \, {}^1P^o_1$, which is only 258 cm$^{-1}$ deeper than $5d7s
\, {}^3D_1$ state, provides the main contribution to the PNC
amplitude. To verify their conclusion, we have carried out the
accurate calculation of this amplitude with combined method of
configuration interaction and many-body perturbation
theory~\cite{DzuFlaKoz96b}. Our determination resulted in the PNC
amplitude 30 times smaller than  the
prediction~\cite{XiaMouYou90}. The main reason for the discrepancy
is the strong interaction of the configurations forming $5d7s
\,^3\!D_1$ and $6s7p \, {}^1\!P^o_1$ states which was not
accounted for in Ref.~\cite{XiaMouYou90}. This configuration
interaction leads to significant cancellations of different
contributions to the matrix element $\langle 5d7s \, {}^3\!D_1
|H_W| 6s7p \, {}^1P^o_1 \rangle$ and decreases the contribution to
the PNC amplitude by an order of magnitude.

To reiterate, with the new data from experiments with antiprotonic
atoms~\cite{TrzJasLub01} we reevaluated the role of nuclear-structure uncertainties
in the interpretation of atomic parity violation  with  chains of
isotopes of the same element. We find that the nuclear-structure
uncertainty is reduced by these new data. We compared possible
constraints on the direct ``new physics'' with the most accurate
to date single-isotope probe of parity violation in Cs atom. We
conclude that presently isotopic chain experiments with atoms
having $Z \lesssim 50$ may be competitive with this single-isotope
determination. As the neutron distribution measurements become
more refined (see, e.g., Ref.~\cite{HorPolSou01}) we expect that
competitive probes of parity violation with isotopic ratios of the
same element may become feasible for heavier atoms.

We would like to thank E.N. Fortson, S.J. Pollock, R. Phaneuf, D. Budker, and
M. Kozlov for useful discussions and E. Emmons for comments on the manuscript.
This work was partially
supported by the National Science Foundation.

%\bibliographystyle{revtex}
%\bibliography{pnc,general,Por_ref,mypub,exact,He,hfs,qed,isonotes,twoval,iso_new}

\end{document}